\documentclass{Interspeech2024}
\interspeechcameraready 

\usepackage{cite}
\usepackage{amsmath,amssymb,amsfonts}
\usepackage{graphicx}
\usepackage{textcomp}
\usepackage{xcolor}
\usepackage{graphicx}
\usepackage{multirow}
\usepackage{multicol}
\usepackage{float}
\usepackage{algorithm}
\usepackage{algpseudocode}
\usepackage{array}
\usepackage[export]{adjustbox}

\usepackage{tabularx}
\usepackage{amsfonts}

\usepackage{hyperref}
\usepackage{caption}
\usepackage{subcaption}

\title{Towards Robust Few-shot Class Incremental Learning in Audio Classification using Contrastive Representation}
\name[affiliation={1}]{Riyansha}{Singh} 
\name[affiliation={2}]{Parinita}{Nema}
\name[affiliation={2}]{Vinod K}{Kurmi}
\vspace{-5em}
\email{riyansha@cse.iitk.ac.in, parinita22@iiserb.ac.in, vinodkk@iiserb.ac.in}

\address{$^1$IIT Kanpur, India\thanks{This research was conducted at IISER Bhopal, India}
\\
  $^2$IISER Bhopal, India }

\begin{document}

\maketitle
\vspace{-3em}
\begin{abstract}
   In machine learning applications, gradual data ingress is common, especially in audio processing where incremental learning is vital for real-time analytics. Few-shot class-incremental learning addresses challenges arising from limited incoming data. Existing methods often integrate additional trainable components or rely on a fixed embedding extractor post-training on base sessions to mitigate concerns related to catastrophic forgetting and the dangers of model overfitting. However, using cross-entropy loss alone during base session training is suboptimal for audio data. To address this, we propose incorporating supervised contrastive learning to refine the representation space, enhancing discriminative power and leading to better generalization since it facilitates seamless integration of incremental classes, upon arrival. Experimental results on NSynth and LibriSpeech datasets with 100 classes, as well as ESC dataset with 50 and 10 classes, demonstrate state-of-the-art performance. Project page: \href{https://visdomlab.github.io/FsACLearning/}{\texttt{https://visdomlab.github.io/FsACLearning/}}.
\end{abstract}

\begin{figure*}[h]
  \centering
  \includegraphics[height=0.36\textwidth]{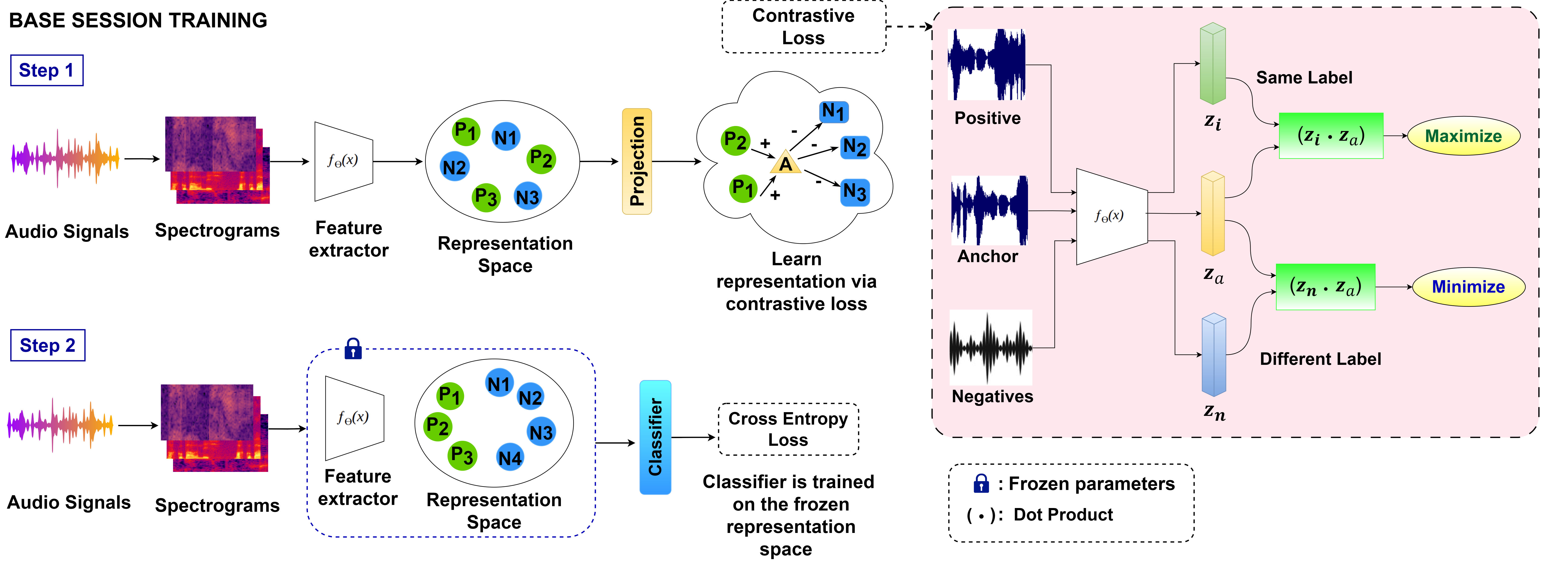} 
  \vspace{-1.1em}
  \caption{During base training, we follow a two-stage process. Initially, the model undergoes training employing the contrastive loss function to seperate classes with minimum overlap. Subsequently, the model is trained utilizing the cross-entropy loss function to guide the optimization process towards better classification performance.}
  \vspace{-2em}
  \label{fig1}
\end{figure*}

\section{Introduction}
Audio classification involves categorizing audio signals into distinct classes like music, speech, or environmental sounds. This task presents unique challenges compared to image-based methods, and many recent efforts have been made in the field ~\cite{nanni2021ensemble, castelbranco2021puremic, kobayashi2014acoustic}. However, existing approaches often assume a fixed audio class vocabulary, and retraining models for new classes is time-consuming and labor-intensive, risking catastrophic forgetting.
Few-shot class-incremental learning (FSCIL) \cite{Tao_2020_CVPR} was introduced as a challenging setting, merging few-shot learning (FSL) and class-incremental learning (CIL). In FSCIL, an initial model is established through a base session, featuring numerous classes and examples. Subsequently, a series of learning sessions follow, where the model progressively learns new classes. Each session involves a unique set of classes sampled from the same distribution, with each class having only a few labeled samples. The challenge lies in ensuring that, after learning new classes, the model retains the ability to recognize all encountered classes. Since we focus on the problem of audio classification, we refer to the problem in the FSCIL setting as few-shot class-incremental audio classification (FCAC) for clarity.

In class-incremental learning, the challenge is balancing stability and plasticity to avoid overfitting on limited new data. Previous works on FCAC reveal bias risk in classes learned from few samples, harming generalization and leading to potential degradation of the base session. In applications where the vocabulary of audio classes dynamically expands,  like in smart audio devices, users often seek new class additions. Few-shot sound classification enables recognizing new classes with minimal samples but struggles with old class retention. Incremental learning approaches aim to address this through a decoupled learning framework with a well-trained initial system comprising a embedding extractor and a classifier. Wang et al.\cite{wang2021few} used dynamical few-shot learning (DFSL) \cite{gidaris2018dynamic} approach by incorporating an attention-based weight generator (ABWG) to expand the  cosine-similarity based classifier of the model. Later Xie et al. proposed an audio classification method via adaptively refined prototypes (ARP) where the prototypes are adaptively refined by a dynamic relation projection module \cite{xie2023fewshot}. To ensure the model's flexibility in handling both new and familiar audio classes, it is crucial to expand the classifier component with updated classification prototypes. This expansion involves integrating newly initialized prototypes specifically designed for recently introduced audio classes. Research conducted within this framework has employed established methodologies such as the finetune \cite{dhar2019learning} method and incremental classifier and representation learning (iCaRL), as proposed by Rebuffi et al. \cite{rebuffi2017icarl}. A more recent advancement employs a stochastic classifier (SC) approach demonstrated by Li et al. \cite{li2023few}. This method systematically expands the classifier during incremental sessions, enabling continual recognition of new classes while keeping the embedding extractor frozen. In contrast, mainstream FCAC frameworks typically rely solely on cross-entropy (CE) during base session training, followed by freezing the backbone to adapt to new classes. However, CE has known limitations, including vulnerability to noisy labels and potential for poor margins, leading to reduced generalization performance. Hence what defines a good base session-model capable of effective generalization to new classes with limited data?

Intuitively, a good base session-model should have representations of base classes concentrated around their clustering centers (prototypes), with prototypes well-separated from each other. This separation facilitates the incorporation of novel classes into the existing space without overlap. Thus, we propose novel methods with improved clustering effects, that is supervised contrastive learning where embeddings from the same class are pulled closer together, enhancing base class separation and, consequently, improving generalization to new classes. \\ 
The contribution of this paper is summarized as follows:

\begin{itemize}
  \item Our novel framework utilizes supervised contrastive loss to train base session to enhance the capabilities of accommodating future classes efficiently.
  \item  We employ a stochastic classifier instead of a deterministic one, dynamically generating prototypes based on learned distributions to continually expand the classifier for new classes. 

  \item We provide a thorough evaluation of the method by
considering detailed comparison on standard benchmark datasets against the state of the art methods.
\end{itemize}


\section{Methodology}

\subsection{Problem Statement}

In the FCAC setup, the learner receives a limited number of labeled audio samples per class in each incremental session. The model's objective is to recognize new audio classes while retaining knowledge of old ones. Incremental sessions occur sequentially, with access to training samples from previous sessions no longer available in each subsequent session.

The training and evaluation datasets of base and ${M}$ incremental session are denoted as \{$P^0_{tr}$, $P^1_{tr}$, .., $P^m_{tr}$, .., $P^M_{tr}$\} and \{$P^0_{ev}$, $P^1_{ev}$, .., $P^m_{tr}$, ..,$P^M_{ev}$\} respectively where class labels for the ${i}$-th session; that is, $P^i$ is represented by $C_i$. The datasets in different sessions do not have the same kind of audio classes, i.e for all i, j, where $i\neq j$,     $C_i\cap C_j = \phi$. The model is assessed using evaluation data from the current session and classes seen in all previous sessions, i.e., $C_0 \cap C_1 ...\cap C_i$, after its incremental session. According to FSCIL, it is conventional to have way more training data in the base session (represented as session 0) $P^0_{tr}$ than in any incremental session $P^1_{tr}$ to $P^M_{tr}$. Each of ${m}$ incremental session consists of ${N}$ audio classes, and each audio class has ${K}$ training samples. Therefore, the training dataset $P^{i}_{\text{tr}}$ in the $i$-th incremental session ($1 \leq i \leq M$) is projected as an $N$-way $K$-shot training problem. 
\vspace{-0.8em}
\subsection{Proposed Model}
The framework of our method is summarized in Figure \ref{fig1}. The architecture can be decomposed into a embedding extractor and the classifier. Taking inspiration from the notable success of ResNet ~\cite{he2016deep} in extracting embeddings for audio and visual processing tasks ~\cite{Shi2022coo, Qian2021Sp}, we have employed the backbone architecture of ResNet18 as the embedding extractor module in our current approach. The input for the embedding extractor consists of log-mel spectrogram, a prevalent choice for neural network input in embedding learning tasks. The extraction process involves several steps: initially, each audio sample undergoes segmentation into overlapping frames of fixed duration, which are subsequently subjected to windowing using a hamming window. Following this, a fast fourier transform is applied to these windowed frames to yield power spectra, which are then convolved with a series of mel-scale filters for spectral smoothing. Finally, the logarithm operation is applied to the resulting Mel spectrum outputs, yielding the log-mel spectrogram.

\noindent A number of provident research work suggest an incremental-frozen framework that optimizes the loss to train the model by utilizing enough data from the base session, that is, the softmax layer which serves the purpose of training the embedding extractor is subsequently excluded from further network operations after the initial training phase. The model $\phi(x)$ is a composition of a embedding extractor and a classifier. Let $\phi(x)$ represent the output vector in $\mathbb{R}^{|C_0| \times 1}$, $W$ denote the weight of classifier which can be represented as $\mathbb{R}^{d \times |C_0|}$ matrix initially where average embeddings belonging to the same class are used to initialize $W$ and $f_\Theta(\mathbf{x})$ is the embedding vector in $\mathbb{R}^{d \times 1}$. This can be articulated as:
 \vspace{-0.6em}
 \small{\begin{equation}
                \vspace{-0.7em}
                \phi(x) = W^{T}f_\Theta(\mathbf{x}) \label{first}
                \vspace{-0.3em}    
            \end{equation}}
        
where $\Theta$ is the parameter vector for the embedding extractor model $f$. In the incremental training phase, the classifier's backbone is fully fixed, and each novel session will see its extension: $W$ is equivalent to \{$w_1^{0}, w_2^{0},...,w_{|C_0|}^{0}$\} $\cup....\cup$ \{$w_1^{t},...,w_{|C_t|}^{t}$\} where $w^{i}$ represents the prototype vector for samples encountered in $i$-th training session. For the purpose of introducing variability in the classification, stochastic classifier is used instead of deterministic where the weight matrix $W$ is given by
\vspace{-0.5em}
\begin{equation}
             W=\mu+\mathcal{N}(0,1) \odot \sigma \label{second}
             \vspace{-0.3em}
\end{equation}
Here the mean vector $\mu$, consisting of average embeddings of all encountered classes is perturbed by adding a certain amount of noise or variability $\sigma$  sampled from $\mathcal{N}(0,1)$ where $\sigma$ is $\mathbb{R}^{d \times |C|}$ variance vector of all classes encountered till $i$-th training session. This technique helps prevent overfitting by introducing randomness during training, which can lead to a more robust model that generalizes better to unseen data. 

\subsection{Enhancing feature representation space using contrastive learning}
\subsubsection{\textbf{Base session training}}
The incremental-frozen baseline, highlighted in studies~\cite{Guan2021, zhang2021few}, outperforms various trainable methods in FSCIL, emphasizing the importance of preserving old knowledge. This highlights the importance of maintaining base performance through the freezing strategy. In contemplating what the model should learn during base training for future few-shot generalization, an intuitive hypothesis suggests that enhancing base class separation contributes to effective novel class generalization. The idea is that if prototypes are widely spaced and class features concentrate around them, it provides ample room for incorporating future classes without overlap. Recognizing the limitations of CE in generalization due to poor class margins~\cite{Gamale2018, Weyang2016}, we explore SCL as an alternative, leveraging its demonstrated clustering effect for improved performance for FCAC~\cite{li2023few} setting which can be very well observed from Figure~\ref{fig2}. 

We utilize the SupCon framework~\cite{khosla2020supervised} method for contrastive training where multiple positives and negatives are considered, deviating from conventional data augmentation for the anchor. For a given input data batch, first we forward propagate it through the encoder, producing a d-dimensional normalized embedding. Contrastive training involves two steps: Initially, SCL promotes representation learning by encouraging closeness among positive pairs and significant separation from anchors for negative pairs, aiming for effective base class separation. Subsequently, in second step after freezing model parameters, the classifier is trained using CE, with embeddings from the same class forming positive pairs and features from different classes forming negative pairs. The objective is to keep anchors close to positive samples and far from negative samples. Within a training batch, let $a \in A \equiv \{1 \ldots N\}$
 be the index of an arbitrary  sample. To this end, given an audio log mel-spectrogram, per-sample SCL is defined as:
 \vspace{-1.5em}

            \begin{equation}
                \mathcal{L}_{\text {CL}}=\sum_{a \in A} \frac{-1}{|I(a)|} \sum_{i \in I(a)} \log \frac{\exp \left(\boldsymbol{z}_a \cdot \boldsymbol{z}_i / \tau\right)}{\sum_{n \in N(a)} \exp \left(\boldsymbol{z}_a \cdot \boldsymbol{z}_n / \tau\right)}\label{third}
                \vspace{-0.13em}
            \end{equation}
            
We consider $\boldsymbol{z}_a$ as the feature embedding extracted for the anchor instance i.e $ \boldsymbol{z}_a = f_\Theta(\mathbf{x_a})$. Similarly, $\boldsymbol{z}_i$ represent the extracted feature embeddings of positive sample that is, sample belonging to same class as $\boldsymbol{z}_a$. For any sample $a$, $I(a) = \{i \in N(a) \mid y_p = y_a\}$ which represents the set of indices for all samples that share the same label as the anchor sample $a$ and $N(a) = A \setminus \{a\}$ represents the set with all indices other than $a$. The temperature parameter is denoted by $\tau$, and $|I(a)|$ denotes the cardinality of set $I$. $(\cdot)$ denotes the dot product. 

\noindent During training, a projection network processes this representation for SCL computation, while for classification at inference, a linear classifier is trained on the frozen representations using CE which is defined as:
\vspace{-0.8em}
\begin{equation}
    \mathcal{L}_{\text{CE}}(\mathbf{x}, y; \Theta) = -\log\left(\frac{e^{\text{cos}(f_\Theta(\mathbf{x}), w_y)}}{\sum_{h=0}^{|\textbf{C}|} e^{\text{cos}(f_\Theta(\mathbf{x}), w_h)}}\right)\label{fourth}
\end{equation}

Here $|\textbf{C}|$ = $C_0 \cup C_1\cup...\cup C_m$. For a given sample x, cosine similarity, $\mathbf{cos}(f_\Theta(\mathbf{x}), W)$ is used to compute the class score for a given feature vector which is defined as:
\vspace{-0.5em}
{\small
\begin{equation}
    \mathbf{cos}(f_{\Theta}(\mathbf{x}), W) = \frac{f_\Theta(\mathbf{x}) \cdot W}{\ ||f_\Theta(\mathbf{x})\ || \cdot \ ||W\ ||} \label{fifth},
\end{equation}
 \vspace{-0.7em}
}

where $||.||$ denotes 2-norm.
Therefore the model is episodically trained on each batch using the joint loss given by
\vspace{-0.3em}
          \begin{equation}
              \mathcal{L} = \lambda \mathcal L_\text{CE} + \beta \mathcal L_\text{CL}\label{sixth},
             \end{equation}

where $\lambda$ and $\beta$ are adjustable coefficients. $\mathcal{L}_{CE}$ and $\mathcal{L}_{CL}$ denotes cross-entropy and contrastive loss function respectively.

\subsubsection{\textbf{Incremental session training}}

For the $m^\text{th}$ incremental session, features of the support set in $P_{tr}^m$  is first extracted. Subsequently, we train the stochastic classifier ~\cite{li2023few} using embeddings from the samples in current session and prototypes from all previous sessions. The prototype loss $\mathcal{L}_p$ is defined by

\vspace{-1em}
\begin{equation}
     \mathcal{L}_p(p_{c},c) = -\log\left(\frac{e^{cos(p_c,{w}_c)}}{\sum_{h=0}^{|\textbf{C}|} e^{cos(p_h,{w}_h)}}\right)\label{eight},
 \end{equation}
 
 where $c \in (C_0 \cup C_1 \cup...\cup C_{m-1})$ and $p_c$ is the prototype vector of class c.  The per session training incorporates a joint loss $\mathcal{L}$ given by a combination of the primary loss $\mathcal{L}_{CE}$ defined earlier and prototype loss $\mathcal{L}_{p}$ defined as:
 \vspace{-0.3em}
 \begin{equation}
     \mathcal{L} = \alpha\mathcal L_{p} + (1-\alpha)\mathcal L_\text{CE} \label{ninth}
     \vspace{-0.5em}
             \end{equation}
            
The adjustable coefficient $\alpha$ balances these two losses. The mean vectors of the trained classifier are used as updated prototype for evaluation.

\begin{table}[!]
\caption{Result obtained by various methods on LS-100 Dataset. "Base," "Incr.," and "All" denotes the performance (accuracy) for base, incremental, and combined sessions.}
\vspace{-2.2em}
\label{tab3}
\begin{center}
\scalebox{0.65}{
\begin{tabular}{p{0.85cm}l|p{0.5cm}p{0.5cm}p{0.5cm}p{0.5cm}p{0.5cm}p{0.5cm}p{0.5cm}p{0.5cm}p{0.5cm}p{0.5cm}p{0.5cm}}
\hline
\multirow{2}{*}{Methods}  & \multicolumn{10}{c}{Accuracy in various sessions (\%)}                                                    & \begin{tabular}[c]{@{}l@{}}AA($\uparrow$)\\ (\%)\end{tabular} & \begin{tabular}[c]{@{}l@{}}PD($\downarrow$)\\ (\%)\end{tabular} \\ \cline{2-13} 
                          & Sess. & 0 & 1 & 2 & 3 & 4 & 5 & 6 & 7 & 8 &                                                   &                                                   \\ \hline
\multirow{3}{*}{FT\cite{dhar2019learning}} & Base    & 92.02   & 72.90     & 37.03    & 28.12    & 20.75    & 14.45    & 5.70      & 3.23     & 0.27     & 30.50                                              & 91.75                                             \\ 
                          & Incr.   & -       & 86.60     & 31.50     & 28.87    & 25.45    & 24.24    & 18.17    & 13.46    & 11.80     & 30.01                                             & 74.80                                              \\ 
                          & All     & 92.02   & 73.95    & 36.24    & 28.27    & 21.93    & 17.33    & 9.86     & 7.00        & 4.88     & 32.39                                             & 87.14                                             \\ \hline
\multirow{3}{*}{iCaRL\cite{rebuffi2017icarl}}    & Base    & 92.02   & 80.80     & 73.18    & 58.45    & 26.95    & 16.93    & 32.58    & 29.53    & 26.38    & 48.54                                             & 65.64                                             \\  
                          & Incr.   & -       & 58.00       & 67.10     & 57.40     & 20.05    & 16.48    & 30.33    & 26.83    & 28.95    & 38.14                                             & 29.05                                             \\ 
                          & All     & 92.02   & 79.05    & 72.31    & 58.24    & 25.23    & 16.80     & 31.83    & 28.54    & 27.41    & 47.94                                             & 64.61                                             \\ \hline
\multirow{3}{*}{DFSL~\cite{wang2021few}}     & Base    & 91.93   & 91.93    & 91.88    & 91.85    & 91.83    & 91.86    & 91.85    & 91.85    & 91.84    & 91.87                                             & 0.09                                              \\ 
                          & Incr.   & -       & 53.60     & 61.90     & 50.67    & 48.90     & 51.56    & 47.97    & 44.11    & 45.38    & 50.51                                             & 8.22                                              \\ 
                          & All     & 91.93   & 88.97    & 87.60     & 83.61    & 81.11    & 80.01    & 77.22    & 74.26    & 73.25    & 81.99                                             & 18.68                                             \\ \hline
\multirow{3}{*}{CEC~\cite{zhang2021few}}      & Base    & 91.72   & 91.67    & 91.25    & 91.14    & 91.10     & 91.07    & 90.97    & 90.66    & 90.72    & 91.14                                             & 1.00                                                 \\ 
                          & Incr.   & -       & 86.30     & 82.76    & 69.67    & 68.25    & 67.06    & 66.03    & 60.35    & 60.05    & 70.06                                             & 26.25                                             \\ 
                          & All     & 91.72   & 91.25    & 90.04    & 86.84    & 85.38    & 84.01    & 82.65    & 79.49    & 78.45    & 85.54                                             & 13.27                                             \\ \hline
\multirow{3}{*}{SC~\cite{li2023few}}     & Base    & 92.73   & 92.72    & 92.62    & 92.48    & 92.48    & 92.47    & 92.34    & 90.74    & 90.67    & 92.14                                             & 2.06                                              \\ 
                          & Incr.   & -       & 86.84    & 84.26    & 77.74    & 74.99    & 75.79    & 74.60     & 72.45    & 72.64    & 77.41                                             & 14.20                                              \\  
                          & All     & 92.73   & 92.27    & 91.42    & 89.53    & 88.10     & 87.56    & 86.43    & 84.00       & 83.45    & 88.39                                             & 9.28                                              \\ \hline
\multirow{3}{*}{Ours}     & Base    & 92.97   & 92.80     & 92.37    & 91.50     & 91.58    & 91.90     & 91.70     & 91.03    & 90.88    & 91.86                                             & 2.09                                              \\ 
                          & Incr.   & -       & 99.60     & 97.00       & 92.73    & 91.05    & 89.64    & 89.43    & 86.14    & 85.63    & 91.40                                              & 13.97                                             \\  
                          & All    & \textbf{92.97}   & \textbf{93.32}    &\textbf{93.03}    & \textbf{91.75}    &\textbf{91.46}    & \textbf{91.24}    &\textbf{90.95}    & \textbf{89.23}    &\textbf{88.78}    & \textbf{91.41}                                             & \textbf{4.18}                                              \\ \hline
\end{tabular}
}
\end{center}
\vspace{-3.5em}
\end{table}

\begin{table*}[h!]
\centering
\caption{Result obtained by various method on NSynth-100 Dataset}
\vspace{-1em}
\label{tab2}
\scalebox{0.75}{
\begin{tabular}{llllllllllllll}
\hline
\multirow{2}{*}{Methods}  & \multicolumn{11}{c}{Accuracy in various sessions (\%)}                                                               & \begin{tabular}[c]{@{}l@{}}AA ($\uparrow$)\\ (\%)\end{tabular} & \begin{tabular}[c]{@{}l@{}}PD($\downarrow$)\\ (\%)\end{tabular} \\ \cline{2-14} 
                          & Session & 0 & 1 & 2 & 3 & 4 & 5 & 6 & 7 & 8 & 9 &                                                   &                                                   \\ \hline
\multirow{3}{*}{Finetune ~\cite{dhar2019learning}} & Base    & 99.96   & 88.91    & 85.41    & 80.36    & 72.51    & 45.24    & 59.31    & 48.53    & 50.68    & 53.28    & 68.42                                             & 46.68                                             \\ 
                          & Incr.   & -       & 38.75    & 30.25    & 36.96    & 37.54    & 28.95    & 27.24    & 22.30     & 20.58    & 19.00       & 29.06                                             & 19.75                                             \\  
                          & All     & 99.96   & 84.73    & 76.92    & 71.06    & 63.18    & 40.15    & 47.99    & 38.33    & 38.01    & 37.86    & 59.82                                             & 62.10                                              \\ \hline
\multirow{3}{*}{iCaRL~\cite{rebuffi2017icarl}}    & Base    & 99.98   & 98.42    & 99.25    & 98.40     & 94.56    & 82.36    & 85.09    & 80.59    & 75.78    & 74.53    & 88.90                                              & 25.45                                             \\  
                          & Incr.   & -       & 36.94    & 31.88    & 35.03    & 38.33    & 35.27    & 30.76    & 26.75    & 25.52    & 22.27    & 31.42                                             & 14.67                                             \\  
                          & All     & 99.98   & 93.30     & 88.88    & 84.82    & 79.57    & 67.65    & 65.92    & 59.65    & 54.62    & 51.01    & 74.54                                             & 48.97                                             \\ \hline
\multirow{3}{*}{DFSL~\cite{wang2021few}}     & Base    & 99.93   & 99.11    & 98.83    & 95.83    & 94.84    & 94.81    & 94.39    & 93.76    & 92.06    & 91.61    & 95.52                                             & 8.32                                              \\  
                          & Incr.   & -       & 57.01    & 55.57    & 59.89    & 59.35    & 56.46    & 52.29    & 50.94    & 52.57    & 52.49    & 55.17                                             & 4.52                                              \\  
                          & All     & 99.93   & 96.00      & 92.95    & 89.26    & 86.47    & 83.66    & 80.28    & 77.68    & 76.12    & 75.01    & 85.74                                             & 24.92                                             \\ \hline
\multirow{3}{*}{CEC~\cite{zhang2021few}}      & Base    & 99.96   & 99.87    & 99.90     & 99.29    & 99.24    & 99.30     & 99.26    & 99.24    & 99.20     & 99.23    & 99.45                                             & 0.73                                              \\ 
                          & Incr.   & -       & 71.06    & 71.61    & 72.37    & 69.17    & 69.20     & 66.92    & 64.80     & 65.28    & 63.59    & 68.22                                             & 7.47                                              \\  
                          & All     & 99.96   & 97.47    & 95.56    & 93.52    & 91.22    & 89.90     & 87.85    & 85.84    & 84.92    & 83.19    & 90.94                                             & 16.77                                             \\ \hline
\multirow{3}{*}{SC~\cite{li2023few}}     & Base    & 99.98   & 98.08    & 98.69    & 97.38    & 96.44    & 97.43    & 96.99    & 97.53    & 96.10     & 96.81    & 97.56                                             & 3.17                                              \\ 
                          & Incr.   & -       & 95.60     & 94.73    & 93.45    & 92.53    & 85.20     & 81.53    & 78.50     & 79.44    & 77.86    & 86.53                                             & 17.74                                             \\ 
                          & All     & 99.98   & 97.88    & 98.08    & 96.53    & 95.55    & 93.61    & 91.54    & 90.13    & 89.09    & 88.29    & 94.07                                             & 11.69                                             \\ \hline
\multirow{3}{*}{Ours}     & Base    & 100     & 99.66    & 99.84    & 98.53    & 98.40     & 98.70     & 98.02    & 97.82    & 98.13    & 97.22    & 98.63                                             & 2.78                                              \\ 
                          & Incr.   & -       & 96.60     & 93.60     & 90.93    & 90.65    & 88.64    & 85.17    & 84.06    & 85.68    & 85.02    & 88.93                                             & 11.58                                             \\ 
                          & All    & \textbf{100}    & \textbf {99.40}     & \textbf{98.88}    & \textbf{96.90}     &\textbf{96.33}    & \textbf{95.55}    &\textbf{93.50}     & \textbf{92.47}    &\textbf{92.88}    & \textbf{91.73}    & \textbf{95.77 }                                           & \textbf{8.27}                                             \\ \hline
\end{tabular}
}
\vspace{-0.5em}
\end{table*}

\vspace{-0.5em}
\section{Results and Experiments}
\textbf{Datasets: }The experiments encompass two distinct audio corpora: NSynth~\cite{engel2017neural} and LibriSpeech~\cite{panayotov2015librispeech}. NSynth comprises 306,043 4 sec audio snippets, each representing one of 1,006 musical instruments. LibriSpeech, a speech corpus, contains around 1,000 hours of audiobooks from 2,484 speakers. For experimental purposes, 100 classes are selected from both LibriSpeech and NSynth datasets referred in the later text as NSynth-100 and LS-100 respectively.
Concurrently, experiments are conducted on prominent environment sound  datasets~\cite{piczak2015} ESC-50 and ESC-10 with 50 and 10 classes respectively, each snippet being 5 sec long.  These datasets are partitioned into two non-overlapping segments, denoted as $P^0$ and $P^t (1 \leq t \leq M)$. $P^0$(or $P^t$) comprises the training dataset $P^0_{tr}$ (or $P^t_{tr}$) and the testing dataset $P^0_{ev}$ (or $P^t_{ev}$ ). Regarding the split of datasets, NSynth-100 has 45 novel and 55 base classes, LS-100 has 60 base and 40 novel classes (both 5-way 5-shot), ESC-50 has 32 base and 18 novel classes, and ESC-10 has 6 base and 4 incremental classes (both 2-way 5-shot). \\
\textbf{Training: }
We employ ResNet18~\cite{he2016deep} as our model to extract representations. We utilize the SGD optimizer with a momentum of 0.9 and a learning rate of 0.1 for 200 base session epochs and an additional 200 new session epochs. Hyper-parameter values are set with $\lambda = 0.2$ for the cross-entropy loss and  $\beta = 1$ for the contrastive loss, ensuring the controlled impact of these two losses across all datasets. The dimensions of the prototypes are 512, while the mel-spectra consists of 128 dimensions. The frames have a length of 25ms with overlapping frames set at 10ms. In each session, accuracy $\mathcal{A}_l$ is measured. The average accuracy across all sessions, denoted by AA = $\frac{1}{L} \sum_{l=0}^{L-1}\mathcal{A}_l$. Performance dropping rate PD = $\mathcal{A}_0$ - $\mathcal{A}_{L-1}$ is used as performance metrics.  In the context of Incr. session, $\mathcal{A}_1$ represents the accuracy of the initial session such that AA = $\frac{1}{L-1} \sum_{l=1}^{L-1}\mathcal{A}_l$ and PD = $\mathcal{A}_1$ - $\mathcal{A}_{L-1}$.



\subsection{Experimental results}
\begin{figure}[!]
     \centering
     
     \begin{subfigure}[b]{0.49\columnwidth}
         \centering
         \includegraphics[width=\textwidth]{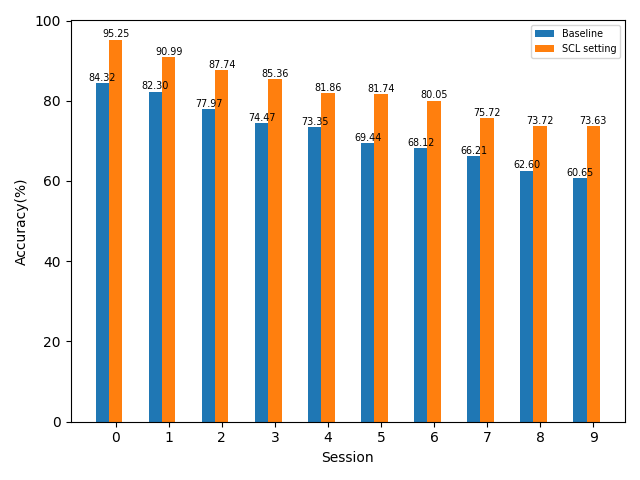}
         \vspace{-1em}
 \caption{Performance on ESC-50}     
     \end{subfigure}
     \hfill
     \begin{subfigure}[b]{0.49\columnwidth}
         \centering
         \includegraphics[width=\textwidth]{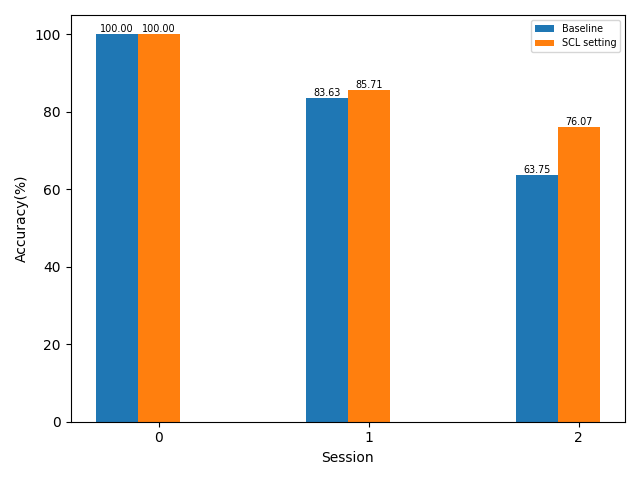}
         \vspace{-1em}
  \caption{Performance on ESC-10}
     \end{subfigure}
     \vspace{-1em}
        \caption{Comparion of per-session accuracy of baseline and our SCL on ESC-50 and ESC-10 trained for 10 and 3 sessions respectively with '0' denoting the base session.}
        \label{fig4}
        \vspace{-2em}
\end{figure}

Our method is compared against existing FSCIL benchmarks, including Finetune~\cite{dhar2019learning}, iCaRL~\cite{rebuffi2017icarl}, DFSL~\cite{wang2021few}, CEC~\cite{zhang2021few}, and SC~\cite{li2023few}. Across all four datasets, we achieve superior performance in terms of accuracy and performance dropping rate (PD). Notably, our method achieves the highest accuracy of 91.41\% for the LS-100 dataset, with a PD rate of 4.18\% (Table \ref{tab3}). Similarly, for the NSynth-100 dataset, we achieve an accuracy of 95.77\% with a PD rate of 8.27\%  (Table \ref{tab2}). Additionally, for the ESC-50 and ESC-10 datasets, our method achieves accuracies of 82.6\% and 87.26, respectively, with corresponding PD rates of 21.5\% and 23.93\% shown in Figure~\ref{fig4}. The visualization of accuracy obtained by our method in last session of ESC-50 is shown in figure~\ref{fig4} (a). It shows the impact of varying values of N and K on method performance. Analysis of accuracy scores,  reveal that optimal performance (88.48\% accuracy) is achieved at (N, K) = (2, 10). Secondly, increasing the number of shots enhances accuracy (except 5-way 5-shot), suggesting improved class understanding with more exposure.  

\subsection{Analysis}
\begin{figure}[!]
    \centering
    \begin{subfigure}[t]{0.49\columnwidth} 
        \centering
        \includegraphics[width=\textwidth, height=5cm, keepaspectratio]{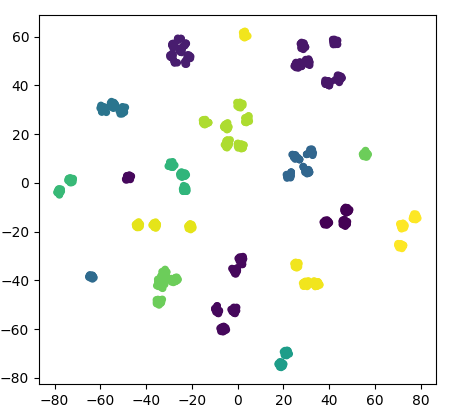} 
        \caption{Stochastic classifier}
    \end{subfigure}
    ~\hfill
    \begin{subfigure}[t]{0.49\columnwidth} 
        \centering
        ~\includegraphics[width=\textwidth, height=5cm, keepaspectratio]{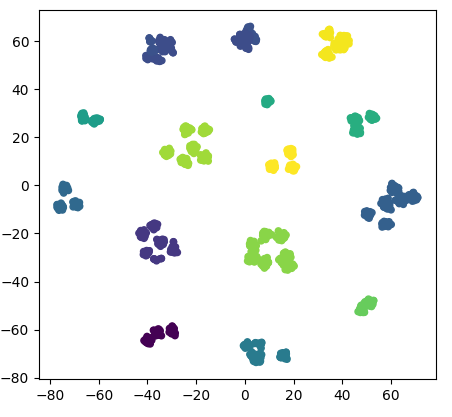} 
        \caption{Proposed model}
    \end{subfigure}
    \vspace{-1em}
    \caption{Base Class Embedding Space representation for LS-100 }
    \label{fig2}
    \vspace{-2.3em}
\end{figure}

\begin{figure}[!h]
     \centering
     \vspace{0.7em}
     
     \begin{subfigure}[t]{0.49\columnwidth}
         \centering
         \includegraphics[width=4.5cm,valign=b]{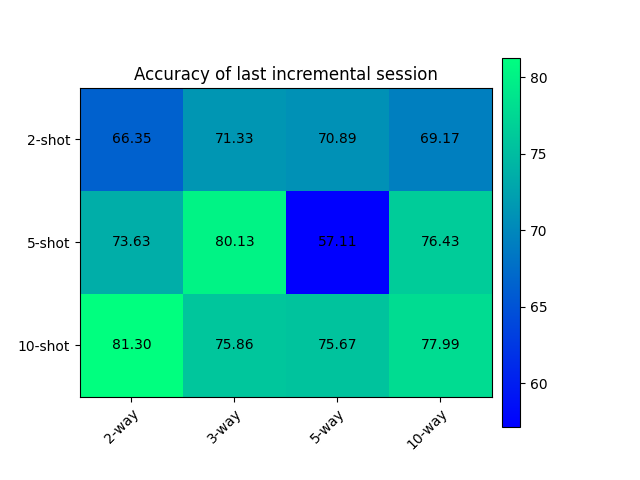}
          
 \caption{Accuracy of the last session obtained by our method on ESC-50}     
     \end{subfigure}
     \hfill
     \begin{subfigure}[t]{0.49\columnwidth}
         \centering
         \includegraphics[width=4cm,valign=b]{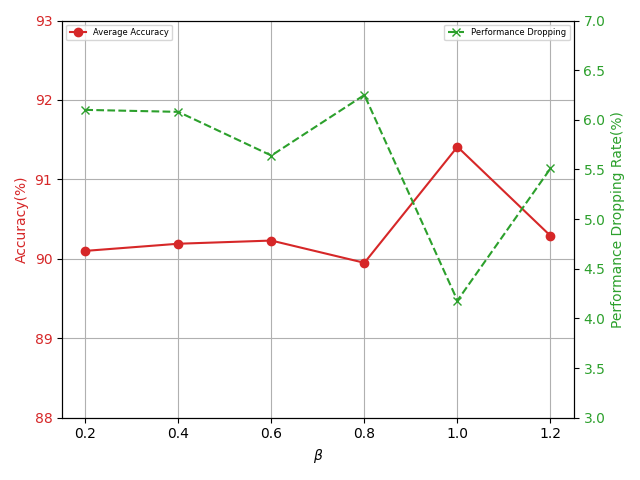}
         
  \caption{Influence of hyperparameter $\beta$ on performance metrics of LS-100 }
  \centering
     \end{subfigure}
     \vspace{-0.5em}
        \caption{(a) shows that a positive correlation exists between the number of shots and the resulting accuracy scores (except 5-way 5-shot) holding the number of ways constant}
        \label{fig3}
\end{figure}

\textbf{Feature visualization:} The embedding space visualization on the LS-100 dataset utilizing t-SNE ~\cite{vandermaaten2008visualizing} is presented in Figure  \ref{fig2}. Fifteen classes from the base classes are randomly selected for this analysis after base training. Our observations reveal that the CE method exhibits limited clustering effects in the embedding space. Our proposed Supervised contrastive learning approach considerably enhances base class separation to seamlessly integrates novel classes with minimal overlap, showcasing its capability to address these challenges. \\
\noindent \textbf{Effect of hyper-parameter of CL:} We employ the hyperparameters $\lambda$ and $\alpha$ to regulate the influence of cross-entropy and contrastive loss, respectively. We explore various values of $\beta$ using the LibriSpeech dataset and Figure \ref{fig3} (b) shows that the optimal performance is achieved when $\lambda = 0.2$ and $\beta = 1$.

\vspace{-6em}

\vspace{3em}


\vspace{1em}

\vspace{-2em}

\section{Conclusion}
In this study, we present a novel framework to address the FCAC challenge which aims to enhance the representation of base classes, facilitating the integration of new classes into the embedding space with minimum overlap. For audio classification, we employ contrastive learning during base session training and use SC for updating prototypes in incremental session. Our method surpasses state-of-the-art approaches in both average accuracy (AA) and performance dropping (PD) across all four datasets. Further improvements are necessary to develop more resilient architectures which are less prone to overfitting during base training, task we entrust to future researchers.


\pagebreak
\newpage
\bibliographystyle{IEEEtran}
\bibliography{main.bbl}

\end{document}